\documentstyle[aps,pra,multicol,psfig]{revtex}
\begin{document}
\draft
\title{The roles of stiffness and excluded volume in DNA denaturation}
\author{Enrico Carlon$^{(1)}$, Enzo Orlandini$^{(1)}$ and
Attilio L. Stella$^{(1,2)}$}
\address{
$^{(1)}$ INFM, Dipartimento di Fisica, Universit\`a di Padova,
I-35131 Padova, Italy \\
$^{(2)}$ Sezione INFN, Universit\`a di Padova, I-35131 Padova, Italy
} 
\date{\today}
\maketitle

\begin{abstract}
The nature and the universal properties of DNA thermal denaturation are
investigated by Monte Carlo simulations. For suitable lattice models we 
determine the exponent $c$ describing the decay of the probability 
distribution of denaturated loops of length $l$, $P \sim l^{-c}$. If 
excluded volume effects are fully taken into account, $c= 2.10(4)$ 
is consistent with a first order transition. The stiffness of 
the double stranded chain has the effect of sharpening the transition, 
if it is continuous, but not of changing its order and the value of the 
exponent $c$, which is also robust with respect to inclusion of specific 
base-pair sequence heterogeneities.  
\end{abstract}

\pacs{PACS numbers: 
87.14.Gg,  
05.70.Jk,  
05.70.Fh,  
87.15.Aa   
}

\begin{multicols}{2} \narrowtext

The melting of a DNA 
molecule from a double stranded phase to a denaturated state where the two 
strands are unbound has been the subject of various studies in the past
\cite{Pola66,Fish66,Peyr89,Cule97,Theo00,Caus00,Kafr00,Gare01}. 
Experiments done with UV absorption by diluted DNA solutions show that 
denaturation occurs through a series of jumps in the absorbance 
spectrum as a function of temperature \cite{Wart85}. 
The jumps are interpreted as sudden denaturations of large double stranded 
portions of the inhomogeneous chain and suggest a first order character for
the transition.
The existing models, even if believed to capture the relevant features
of the system, introduce drastic simplifications of the complex 
structure of the DNA molecule in favor of analytical tractability.
Within these models, two different mechanisms responsible for a first order 
transition were recently proposed. For the Peyrard-Bishop (PB) 
model \cite{Peyr89}, it has been argued that the stronger stiffness 
of double stranded compared to single stranded DNA, may lead to a first order 
denaturation \cite{Theo00}. It has also been proposed that this stiffness 
difference, in combination with base sequence heterogeneity, should be
responsible for the observed jumps in absorption \cite{Cule97}.
Other studies using the Poland-Sheraga (PS) \cite{Pola66} model led to the 
claim that excluded volume effects, even in the absence of stiffness, induce 
first order melting \cite{Kafr00}.

In this Letter we investigate further these issues within models
representing as realistically as possible the relevant properties of DNA.
Within such models all mechanisms mentioned above can operate and thus be 
tested simultaneously without resorting to uncontrollable approximations. 
Near the transition DNA can be regarded as an alternating sequence of
double helix segments along which base pairs are bound, and of denaturated
loops, where the two strands are detached.
By a scaling analysis of the cumulative probability distribution function of 
denaturated loop lengths we show clearly that excluded volume effects drive 
the transition first order, in the limit of infinitely long chains. We also 
find that double helix stiffness alone does not modify such asymptotic 
result in a range of values chosen consistently with those expected for real 
DNA. However, in the presence of strong stiffness, and for a model with
second order denaturation in the infinite chain limit, the critical region 
becomes very narrow and, due to a slow crossover, thermodynamic quantities 
like the specific heat may behave consistently with first order even for long 
finite chains.

We model the DNA strands by two self-avoiding walks (SAW) of length $N$ on 
cubic lattice, identified by the vectors $\vec{r}_1(i)$ and $\vec{r}_2(i)$
($0 \leq i \leq N$), joined at a common origin ($\vec{r}_1(0)= \vec{r}_2(0)$) 
and with free endpoints. 
A gain of energy $\varepsilon$ ($=1$ here) is associated to a bond between 
the two strands, which occurs in the model when two monomers with the same 
$i$ overlap ($\vec{r}_1(i)= \vec{r}_2(i)$). 
The binding energy is taken to be the same all along the chain, i.e., to 
start with, we neglect the heterogeneity of base pair interactions of 
specific sequences.
At sufficiently low T the most probable configurations are fully bound, i.e. 
$\vec{r}_1(i)=\vec{r}_2(i)$ for all $i$. Upon increasing the temperature the 
two strands are expected to unbind at some $T=T_c$. The transition is driven 
by the formation of denaturated loops, whose length can be measured by the 
number of unbound monomers, $l$, of the corresponding strand segments.

\begin{figure}[b]
\centerline{
\psfig{file=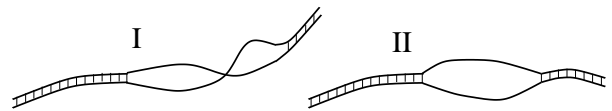,height=1.4cm}}
\vskip 0.15truecm
\caption{Schematic representations of denaturated loops in model I and II.}
\label{FIG00}
\end{figure}  

In order to investigate in detail excluded volume and stiffness effects we 
consider two different models, which we refer to as model I and II. 
While in both models bounded segments are self-avoiding and cannot overlap 
any other part of the chain, in I the excluded volume condition is 
partly relaxed (see Fig. \ref{FIG00}): the two strands forming a denaturated
loop are self-, but not mutually avoiding and overlap at non-complementary
sites (i.e. $\vec{r}_1(j)= \vec{r}_2(k)$ with $j \neq k$) are allowed. Such 
overlaps do not contribute to the energy.
The stiffness of the double stranded chain is incorporated in a second 
parameter $\varepsilon_b$, which is the energy gain if two consecutive 
segments in the bound chain are aligned (i.e. when $\vec{r}_1(i) =
\vec{r}_2(i)$ for $i=j-1,j,j+1$ and $\vec{r}_1(j+1)-\vec{r}_1(j) = 
\vec{r}_1(j)-\vec{r}_1(j-1)$). 
Of course, both models neglect the precise helix structure. Model II, having 
excluded volume effects fully built in, is a more realistic representation of 
DNA. A recent study, in absence of stiffness \cite{Caus00}, claimed for it 
some evidence of first order transition. 

For models I and II we sampled by a multiple Markov chain Monte Carlo (MC)
\cite{Tesi96} method the loop length probability distribution function (pdf) 
at several different temperatures. In all cases the pdf has an exponential
decay at low T ($P(l,N)\sim \exp(-l/l_0)$) where bound segments
are instead broadly distributed in length. From a given $T_c$ upwards one 
observes instead a scaling form $P(l,N) \sim l^{-c} f(l/N)$, where the 
exponent $c$ seems to be approximately independent of $T$ and $f$ is a 
scaling function.
The length of bound double segments is narrow distributed
in this case. The general picture emerging from our numerical results
has an immediate thermodynamical interpretation. Quantities usually analyzed 
at polymer conformational transitions are the energy per monomer or the  
specific heat, which for a chain of length $N$ and $T \sim T_c$ scales as 
\cite{Vand98}:
\begin{equation}
{\cal C}(N,T) \sim N^{2 \phi -1} h\left[(T-T_c) N^\phi \right]
\label{specheat}
\end{equation}
where $h$ is a scaling function and $\phi$ is the crossover exponent.
For large $N$ one has ${\cal C}_{\rm max}(N) = \max_T {\cal C}(N,T) \sim 
N^{2 \phi-1}$, from which $\phi$ can be deduced.
The density of binding contacts along the strands, proportional to the 
energy, should scale as $N^{\phi- 1}$ at $T=T_c$. Clearly 
$\phi \le 1$, and only $\phi=1$ implies a first order discontinuity of 
the density. Given the scaling form of $P(l,N)$ and the fact
that bound segment lengths are finite on average at $T_c$, the same 
density should also scale as the reciprocal of the average loop length
$\langle l \rangle = \sum_l l P(l,N)$. Now, for $1 \leq c < 2$, $\langle l 
\rangle \sim N^{2-c}$, so that $\phi=c-1 <1$, and the transition is second 
order. 
If instead $c > 2$, $\langle l \rangle$ and the energy density remain finite 
at $T_c$ for $N \to \infty$ and the transition is first order ($\phi=1$).
Analyzing $P$ offers both fundamental and practical advantages over the use 
of Eq. (\ref{specheat}). Indeed, due to finite size corrections to
scaling, a reliable estimate of $\phi$ requires good determinations of
${\cal C}$ around $T_c$ for sufficiently long chains \cite{note_causo}. 
On the contrary, the scaling behavior of $P$
sets in already for relatively short chains ($N \approx 100$), 
which allow us to estimate $c$ reliably.
The robustness of the estimate with respect to the variation of chain 
lengths assures that results are little affected by finite size corrections. 

\begin{figure}[b]
\centerline{
\psfig{file=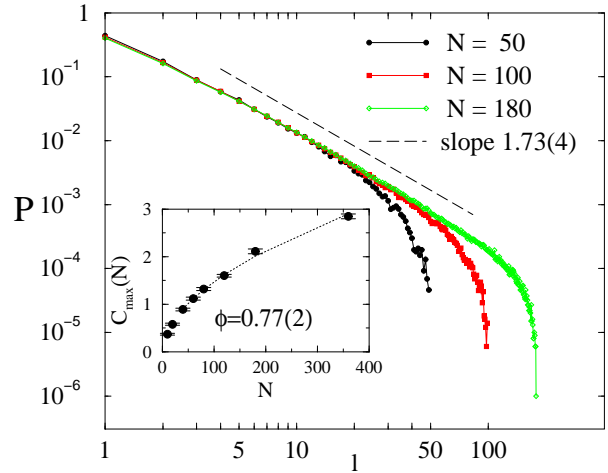,height=6.5cm}}
\vskip 0.15truecm
\caption{Plot of $P$ vs. $l$ on a log-log scale for model I at $T=T_c$ and
various chain lengths $N$. Inset: ${\cal C}_{\rm max}$ as a function of $N$ 
and relative fit $\sim N^{2 \phi -1}$ (see Eq. (\ref{specheat})).}
\label{FIG02}
\end{figure}

We consider first $\varepsilon_b = 0$. Figure \ref{FIG02} shows log-log plots 
of $P$ as a function of $l$ for $N=50,100$ and $180$ for model I, at $T = 
0.85 \approx T_c$ (we estimate $T_c = 0.85(2)$). After an initial transient 
the data follow nicely a straight line in the plots, except when $l \approx N$, 
where of course $P$ drops. It is interesting to note that the power-law 
regime sets in already for relatively short loops. A linear fit of the 
data for $N=180$ gives $c = 1.73(4)$, indicating a continuous denaturation 
for model I. By fitting ${\cal C}_{\rm max}$ as 
${\cal C}_{\rm max}(N) = A N^{2 \phi -1 }$ (inset of Fig. \ref{FIG02}),
we obtain $\phi = 
0.77(2)$. Thus again $\phi < 1$ indicates a continuous transition; moreover, 
the above proposed relation $\phi=c-1$ is well satisfied.

Within a PS model description each loop would be assumed to be totally
uncorrelated with the rest of the chain, which at $T \geq T_c$ would just 
provide a grand-canonical critical reservoir of strand elements to form the 
loop. The value of $c$ within this scheme can be calculated
easily. The critical grand-canonical pdf of a SAW of length $l$ with fixed
end-to-end distance $R_e$, in three dimensions, is \cite{Vand98}:
\begin{equation}
p(l, R_e) \sim  l^{\gamma - 3 \nu -1} g(R_e l^{-\nu}),
\label{part}
\end{equation}
where $g$ is a scaling function, $\gamma$ is the entropic, and $\nu$ is the 
metric exponent. A loop in model I is made of two strands with common 
end-points and allowed to intersect each other. Thus
the probability of a loop of length $l$ (total perimeter $2l$) is:
\begin{equation}
P(l) \sim \int d^3 R_e \,\, p^2(l, R_e) \sim l^{2\gamma - 3 \nu -2}
\label{part2}
\end{equation}  
from which follows $c^{\rm (IL)} = 2 +3 \nu -2 \gamma$, where $\rm{IL}$
stands for isolated loop. Using the appropriate SAW exponents, 
i.e. $\gamma \approx 1.158$, $\nu \approx 0.588$ \cite{Vand98}, we find 
$c^{\rm (IL)} \approx 1.448$ \cite{nota}.
This result allows us to quantify the variation of the exponent $c$ due to
excluded volume interactions between loops and segments which are taken into 
account in the MC simulations, namely $\Delta c = c - c^{\rm (IL)} \approx 
0.3$.

\begin{figure}[b]
\centerline{
\psfig{file=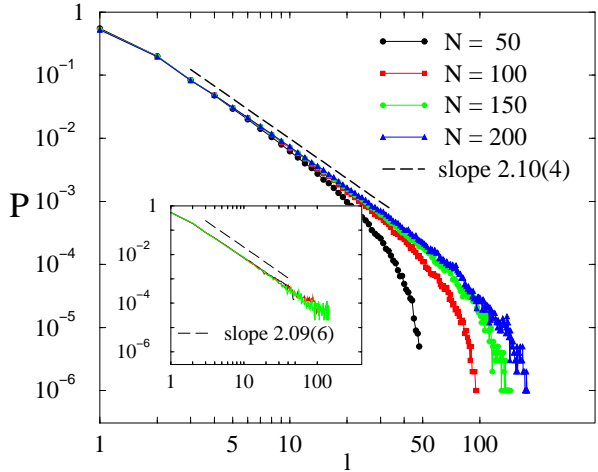,height=6.5cm}}
\vskip 0.05truecm
\caption{Log-log plots of $P$ vs. $l$ for model II at $T=T_c$ and for various 
$N$. Inset: Same plots for $T = 1.3 T_c$.}
\label{FIG03}
\end{figure}

We now turn to model II. Fig. \ref{FIG03} shows the loop probability
distribution at $T \approx T_c$ ( $T_c = 0.76(2)$) for $N=50$, 
$100$, $150$ and $200$. From a linear fit of the data we obtain $c = 2.10(4) 
> 2$.
Excluded volume effects appear to be responsible for the discontinuous
nature of denaturation, in agreement with a recent extension of the PS model
in which the interaction between loops and segments was included in an
approximate way \cite{Kafr00}. By relying on field theoretical
results for polymeric networks \cite{Dupl86}, the authors of Ref. \cite{Kafr00}
estimated analytically the loop length pdf exponent to be $c \approx 2.12$, 
which is remarkably close to our numerical determination, suggesting that 
the sort of perturbative treatment of Ref. \cite{Kafr00} catches the essential 
contribution to the correlations among different loops and segments.
The inset of Fig. \ref{FIG03} shows a plot of $P$ for $T=1.0$, i.e. well
above $T_c$, and $N = 150$; the 
data still show a power-law behavior with an exponent $c = 2.09(5)$, i.e. 
consistent with the value found at $T_c$. 

\vbox{
\begin{table}[tb]
\caption{Summary of exponents found for the two models.}
\vskip 0.05truecm
\label{TABLE01}
\begin{tabular}{c|ccc}
Model  & $c^{\rm (IL)}$ & $c(\varepsilon_b=0)$ & $c(\varepsilon_b=5)$ \\
\hline
I  & 1.448 & 1.73(4)  & 1.70(6)  \\
II & 1.762 & 2.10(4)  & 2.06(6)
\end{tabular}
\end{table}
}

Table \ref{TABLE01} summarizes the results obtained for the two models.
In three dimensions the exponent of an isolated loop is $c^{\rm (IL)} \approx 
1.762$ \cite{Vand98}. For both model I and II the excluded volume interactions 
are responsible for a roughly identical increase of $c$ with respect to its 
``bare'', isolated loop value $c^{\rm (IL)}$. For model I this effect is not
strong enough to cause a first order transition. 

Next we ask whether a sufficiently strong $\varepsilon_b > 0 $ may induce 
first order denaturation. Recently such a possibility has been discussed in 
the context of the PB model~\cite{Cule97,Theo00}. 
This model, like the PS one, takes into account excluded volume effects
very inadequately, since the two strands, while prevented from 
overlapping each other, are in fact not embedded in ordinary three dimensional 
space. Only a sort of longitudinal coordinate along the DNA 
backbone and the distance between strands enter the description.
Within this context it was predicted that the stiffness of bound segments 
sharpens a continuous transition, making it look like first order
for practical purposes~\cite{Cule97}, or even drives it 
strictly first order~\cite{Theo00}. 

\begin{figure}[b]
\centerline{
\psfig{file=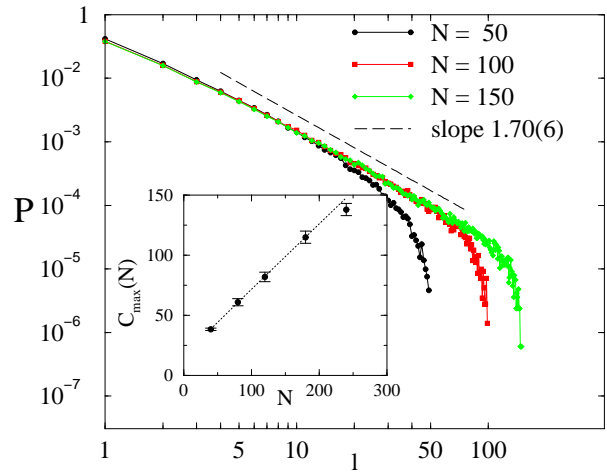,height=6.5cm}}
\vskip 0.15truecm
\caption{Log-log plot of $P$ vs. $l$ for model I with $\varepsilon_b = 5$. 
Inset: scaling of ${\cal C}_{\rm max}$ as a function of $N$.}
\label{FIG04}
\end{figure}

We focus first on model I, which has a continuous transition for 
$\varepsilon_b = 0$. Figure \ref{FIG04} shows a plot of $\ln P$ vs. $\ln l$ 
for $\varepsilon_b=5$.
From a linear fit we obtain $c = 1.70(6)$, in agreement with the value 
determined for $\varepsilon_b=0$. This indicates that a value of stiffness 
$\varepsilon_b=5$
does not change the asymptotic character of the transition, which remains 
second order and in the same universality class as for $\varepsilon_b=0$. 
The inset of Fig. \ref{FIG04} shows a plot of ${\cal C}_{\rm max}$ as 
a function of $N$. This quantity scales linearly 
up to $N \approx 200$, and deviates from linearity only for 
the longest lengths investigated. A linear scaling would imply $\phi = 1$, i.e. 
a first order transition. Apparently there is a very slow crossover in the 
specific heat, i.e. the deviation from ${\cal C}_{\rm max}(N) \sim N$ can 
be observed only for very long chains. Indeed, for $\varepsilon_b=10$,
which can also be a realistic value for DNA, the crossover must occur
at $N$ values which are not accessible to our simulation.
Notice that, on the contrary, already for rather short chains ($N 
\lesssim 100$) the behavior of $P$ indicates clearly that 
$c \sim 1.7 < 2$. 
So, while the investigation of ${\cal C}$ in the presence of
realistic stiffness would reveal the true asymptotic nature
of the transition only for extremely long chains, $c$ seems to
be very little affected by crossover. 
This discrepancy is due to the different way in which finite size 
effects and temperature uncertainties affect the scalings of $P$ and 
${\cal C}$, as discussed above. 

For model II it is not possible to detect crossover behavior in
${\cal C}_{\rm max}$, since the transition is already first order ($\phi=1$) 
at $\varepsilon_b=0$. For $\varepsilon_b=5$
and $\varepsilon_b=10$, which, as discussed below, are realistic choices,
we estimate $c=2.06(6)$ and $c=2.04(8)$, respectively. These values do not 
deviate, within error bars, from that obtained at $\varepsilon_b=0$. However, 
a very slight systematic decrease of $c$ with increasing $\varepsilon_b$ 
cannot be excluded. This decrease could be determined by a mild crossover 
phenomenon induced by $\varepsilon_b > 0$.
Double stranded DNA has a persistence length which is about $20-50$ times 
longer than that of single stranded chains \cite{persist}. According to our 
calculations the persistence lengths of bounded segments for $T \approx 0.8 
T_c$ \cite{note} are about $10$ and $30$ times those of the unbound segments 
for $\varepsilon_b =5$ and $\varepsilon_b = 10$, respectively. Therefore the 
above parameter choices should be considered as rather realistic for DNA.
A single loop attached to very stiff long segments (a typical configuration 
for our chains at low temperatures) should experience very little excluded 
volume interaction with them. However, close to denaturation more and more 
loops start forming and the whole chain becomes rather flexible as the bubbles 
carry no stiffness. Typical estimates in our simulations yield a persistence 
length of just $2-3$ lattice steps close to $T_c$, explaining why the stiffness 
has little effect on the critical loop pdf. For a correct physical 
interpretation of the models studied here one should assign to each lattice 
monomer about $10$ base pairs (bp), thus we predict that the persistence 
length of the double stranded chain close to denaturation would be of the 
order of $20-30$ bp.

We finally considered the effect of heterogeneity of the binding energies of 
base pairs along the chain. As an example we embodied in model II the 
information concerning a specific DNA sequence, resorting to a strictly 
microscopic interpretation of lattice monomers as single bases.
We took $\varepsilon_{\rm CG}/\varepsilon_{\rm AT} = 1.5$ and $\varepsilon_b 
= 5$, as expected for real DNA. For two different sequences of length $N=150$ 
\cite{sequence} we estimated $c = 2.10(8)$, indicating that $c$ robustly 
maintains the value determined for the homogeneous version of the model. 

Another important consequence of our results concerns codes \cite{Blak99}
used to simulate melting curves of real DNA sequences. These codes contain
various parameters, such as the stacking energy of neighboring base pairs, 
and use $c$ in the partition function for denaturated loops. 
The typical choice is $c \approx 1.7$ \cite{Blak99}, which is the appropriate 
value for a single isolated loop, while a more consistent choice would be 
$c \approx 2.1$. It would be interesting to investigate the consequence of 
this different value of $c$ on the calculated melting curves.

Summarizing, the DNA denaturation transition can be characterized in terms 
of the algebraic decay of the cumulative loop length pdf. The exponent $c$ 
signals clearly the asymptotic nature of the transition and is in fact well 
defined already for relatively short chains.
In all our calculations we observe an onset of power-law behavior for short 
loops ($l \gtrsim 5$). This suggests that, if direct measurements could be 
realized, the sampling of $P$ would not need to include very long 
denaturated loops and rather microscopic probes could be adequate.
Promising in this respect seem to be techniques based on fluorescent DNA 
probes (see e.g. Ref. \cite{Bonn99}).

Excluded volume effects alone appear to be responsible for the discontinuous 
nature of denaturation in the infinite chain limit. Our calculations provide 
in fact a final verification of a 
conjecture advanced many decades ago by Fisher~\cite{Fish66} and recently 
corroborated and made more precise by Kafri et al. \cite{Kafr00}.
The stiffness of the double stranded DNA is not responsible for the first 
order character, contrary to claims based on results for the PB model 
\cite{Theo00}, but possibly only produces very mild crossover
effects on $c$. In realistic conditions these effects should be barely
detectable even on relatively short chains. 

Financial supports by MURST-COFIN 01, INFM-PAIS 01 and
European Network  ERBFMRXCT980183 are gratefully acknowledged.

\end{multicols}

\begin{references}
\bibitem{note_causo} The difficulty of a reliable determination of $\phi$ 
from specific heat data has been encountered in Ref.\cite{Caus00}, where the 
estimate $\phi \sim 0.9$, for model II, is not fully consistent with the 
claimed first order nature of the transition.
\bibitem{nota} Notice that using the random walk exponents $\gamma=1$, 
$\nu=1/2$ in the relation $c=2+3 \nu -2 \gamma$ one obtains $c=3/2$, which 
is the correct value of $c$ for a loop where no excluded volume effects 
are considered. It is also interesting to point out that for a loop in 
model I, $c^{\rm (IL)} = 1.448 < 3/2$. This goes against the
expectation that the incorporation of excluded volume effects always leads
to an increase of the value of $c$.
\bibitem{note} We assume a typical denaturation temperature of $T_c \approx
80^o C$, therefore the room temperature, where persistence lengths are 
typically measured, corresponds to $\approx 0.8 T_c$.
\end{references}
\end{document}